\documentclass[twocolumn,showpacs,preprintnumbers,amsmath,amssymb,floatfix]{revtex4}

\usepackage{graphicx}
\usepackage{dcolumn}
\usepackage{bm}
\usepackage{color}

\newcommand{\beq}{\begin{equation}}
\newcommand{\eeq}{\end{equation}}
\newcommand{\bey}{\begin{eqnarray}}
\newcommand{\eey}{\end{eqnarray}}

\begin{document}


\title{Stability analysis of Lower Dimensional Gravastars in noncommutative geometry}

\author{Ayan Banerjee}
 \email{ayan_7575@yahoo.co.in}
\affiliation {Department of Mathematics, Jadavpur University,
 Kolkata-700032, India}

\author{Sudan Hansraj}
 \email{hansrajs@ukzn.ac.za}
\affiliation {Astrophysics and Cosmology Research Unit, School of Mathematics,
Statistics and Computer Science, University of KwaZulu-Natal,
Private Bag X54001, Durban 4000, South Africa}

\date{\today}

\begin{abstract}
The Ba\~{n}ados, Teitelboim and Zanelli \cite{BTZ1992}, black hole solution is revamped
from the Einstein field equations in (2 + 1)-dimensional anti-de Sitter spacetime, in a
context of noncommutative geometry \cite{Rahaman(2013)}. In this article, we explore the exact gravastar solutions
in three-dimension anti-de Sitter space given in the same geometry. As a first step we derive BTZ solution assuming
the source of energy density as point-like structures in favor of smeared objects, where the particle mass M,
is diffused throughout a region of linear size $\sqrt{\alpha}$ and is described by a Gaussian function of finite width
rather than a Dirac delta function. We matched our interior solution to an exterior BTZ spacetime
at a junction interface situated outside the event horizon. Furthermore, stability analysis
is carried out for the specific case when $\chi < 0. 214$ under radial
perturbations about the static equilibrium solutions. To give theoretical support we are also trying 
to explore their physical properties and characteristics.

\end{abstract}

\maketitle

\section{Introduction}

 The recent detection of gravitational waves (\cite{Abbott} and references therein) carried implicit evidence for the existence of black holes since the cataclysmic event originating the wave carried the expected signature of  a coalescing binary black hole system. Nevertheless, the question of what the final fate of gravitational collapse is remains open. Black holes are one possibility but this does not preclude others.
The gravastar (gravitational vacuum star)  model has been proposed by Mazur and Mottola \cite{Mazur,Mottola}, has attracted attention
as an alternative model to black hole.  The general idea is preventing horizon (and singularity) formation, by stopping
the collapse of matter at or near where the event horizon is expected to form i.e., alternative configurations of black holes
could be formed by gravitational collapse of a massive star.  The quasi-normal modes of thin shell non-rotating gravastars were studied by Pani {\it et al} \cite{Pani} and they also considered the gravitational wave signatures when no horizon is present such as in the case of gravastars. To the knowledge of the authors no investigations into the wave signature from coalescing gravastars have been made to date. It has been speculated that the gravastars and black holes emit the same gravitational wave signatures.

In the gravastar  model, the interior
consists of a segment of the de Sitter geometry, enclosed by a shell of Bose-Einstein
condesate, all of which is surrounded by a Schwarzschild vacuum but
without encountering a horizon. The de Sitter interior with negative pressure favouring expansion is necessary to provide a mechanism to counterbalance the gravitational collapse of the ultra-compact Bose-Einstein condensate (BEC) which itself is assumed to have the most extreme equation of state permissible by causality - that of stiff matter ($p = \rho$).
Therefore the gravastar is a multilayered structure consisting of three different regions
with three different equations of state (EOS): \\
(I) an internal core (de Sitter) with an EOS :  p+$\rho$ =0, \\
(II) thin shell of ultra-stiff matter (BEC) with an EOS : p = +$\rho$,\\
(III) outer vacuum Schwarzschild solution with  EOS: p = $\rho$ =0, \\
In practice, the Mazur-Mottola model is a static spherically symmetric with five-layer solution of the Einstein equations
including two infinitesimally-thin shells endowed with surface densities $\sigma_{\pm}$ and surface pressure $ p_{\pm}$.

  Motivated by the work Visser and Wiltshire \cite{Visser}, analysed the dynamic stability against
spherically symmetric perturbations using the Israel thin shell formalism while Carter in \cite{Carter}
has extended  gravastar stability with generalised exteriors (Reissner-Nordstr$\ddot{o}$m). In Ref. \cite {Lobo(2007)}
gravastar solutions have been studied within the context of nonlinear electrodynamics. Later some simplifications and
important remarks about gravastar have been studied in-depth in \cite{Usmani,Horvat,Chan,RChan,Horvat(2007)}. Moreover
the limits on gravastars and how an  external observer can distinguish it from a black hole have been studied in \cite{Broderick,Chirenti}.

After the theoretical discovery of radiating black holes by Hawking \cite{Hawking,Hawking(1975)}, based on
quantum field theory, the thermodynamical properties of  black holes have been studied extensively.
Since this theoretical effort first disclosed the mysteries of quantum gravity, considerable interest in this problem has developed in theoretical physics.  It is generally believed that spacetime as a manifold of points breaks down
at very short distances of the order of the Planck length. In these circumstances noncommutative geometry \cite{Snyder,Seiberg}
plays a key attribute in unraveling the properties of nature at the Planck scale. From the fundamental point of view of  noncommutative geometry there is an interesting interplay  between mathematics, high energy physics as well as cosmology and astrophysics.
In a noncommutative space-time the coordinate operators on a D-brane \cite{Witten} can be encoded by the commutator
$[\hat{x}^{\mu}, \hat{x}^{\nu}]$ = $i\vartheta^{\mu\nu}$, where $\hat{x}$ and $i\vartheta^{\mu\nu}$ are the coordinate
operators and  an antisymmetric tensor of dimension (length)$^2$, which determines the fundamental cell discretization
of spacetime. As Smailagic {\it et al} have shown \cite{Smailagic} that noncommutativity replaces point-like structures
by smeared objects in flat spacetime. Thus it is reasonable to believe that noncommutativity could
eliminate the divergences that normally appear in general relativity that appears in various form.
As discussed in Refs. \cite{Nicolini} The smearing effect is mathematically implemented as a substitution rule :
position Dirac-delta function is substituted everywhere using a Gaussian distribution of minimal
length $\sqrt{\alpha}$.

In the same spirit, Nicolini $\emph{ et al.,}$ \cite{Nicolini,Nicolini(2005),Spallucci(2006)} have investigated
the behavior of a noncommutative radiating Schwarzschild black hole. There is a lot of noncommutative
effects have been performed to extend the solution for higher dimensional black hole \cite{Rizzo},
charged black hole solutions \cite{Ansoldi,Spallucci(2009)} and charged rotating black hole solution
\cite{Smailagic(2010),Modesto}. A number of studies have been performed in these directions where
spacetime is commutative \cite{Oh}. In the same context wormhole solutions have been studied in
\cite{Garattini(2009)}. Recently, Lobo and Garattini \cite{Lobo(2013)} showed that a noncommutative geometry
background is able to account for exact gravastar solutions and studied the linearized stability. Gravastar solutions
in lower dimensional gravity have been studied in \cite{Rahaman(2012)} in an
anti-de Sitter background space-time.

The motivation for this investigation is clear  from the above summary on the aspect of an exact gravastar solution in the context
of NC in (2+1)-dimension. Our paper is organized as follows. In Sec. II we construct BTZ black hole solution
from an exact solution of the Einstein field equations in the context of noncommutative geometry
and specifying the mass function we present the structural equations of gravastar. In Sec. III
we discuss the matching conditions at the junction interface and determine the surface stresses.
In Sec. IV and V we investigate the linearized stability of gravastars and determine the stability
regions of the transition layer. Finally, in Sec. VI we draw the conclusions.

\section{Interior geometry}

We will be concerned here the interior spacetime described by the line element for
a static spherically symmetry and time independent metric in (2+1) dimensions in the
following form :
\begin{equation}
ds^2=-e^{2\Phi(r)}dt^2 + \frac{dr^2}{1-2m(r)/r} +r^2 d\theta^2 ,
\end{equation}
where $\Phi(r)$ and m(r) are arbitrary functions of the radial coordinate, $r$.
Here the "gravity profile" factor $\Phi(r)$ is related with the following relationship:
$\mathcal{A}$ = $\sqrt{1-m(r)/r }\Phi ^{'}(r)$, which represents the locally measured
acceleration due to gravity \cite{Lobo}. The convention used is that $\Phi ^{'}(r)$
is positive or negative for an inwardly gravitational attraction or an outward gravitational
repulsion and m(r) can be interpreted as the mass function.

We take the matter distribution to be anisotropic in nature and therefore the stress-energy
tensor for an anisotropic matter distribution is provided by

\begin{equation}
T_{ij}=(\rho+P_t)u_iu_j+P_tg_{ij}+(p_r-p_{\perp})\mathcal{X}_i\mathcal{X}_j,
\end{equation}
where $u^i$ is the 3-velocity of the fluid and $\mathcal{X}_i$ is the unit spacelike
vector in the radial direction. $\rho(r)$, $p_r(r)$ and $p_{\perp}(r)$ represent the energy density, radial
pressure and tangential pressure, respectively.

The Einstein field equations $G_{\mu\nu}$ +$\Lambda$$g_{\mu\nu}$  = 8$\pi T_{\mu\nu}$, for the spacetime given in Eq. (1)
together with the energy-momentum tensor given in Eq. (2), rendering
G = c = 1, provides the following relationships

\begin{eqnarray}
8\pi\rho+\Lambda &=& \frac{rm'-m}{r^3},\\
8\pi p_r - \Lambda &=& \frac{\Phi^ {'}}{r}\left(1-\frac{2m}{r}\right),\label{eq4}\\
8\pi p_{\perp} - \Lambda &=& \left(1-\frac{2m}{r}\right)\left(\Phi^ {'2}+\Phi^ {''}-\Phi^ {'}\frac{(rm'-m)}{r^2}\right) ,\label{eq5}
\end{eqnarray}
In addition, we have, the conservation equation in (2+1) dimensions:
\begin{equation}
\left(\rho+p_r\right)\Phi^{'}+p_r^{'}+\frac{1}{r}\left(p_r-p_{\perp}\right)=0,
\end{equation}
where $\Lambda$ is the cosmological constant and $\Phi$
are arbitrary functions of the radial coordinate r. Here $`\prime'$ denotes differentiation
with respect to the radial parameter r.  Although we shall not invoke isotropic particle pressure,
it is interesting to note that the isotropy equation $(\ref{eq4}) = (\ref{eq5})$
\beq
\Phi'' + \Phi'^2 - \frac{\Phi'}{r^2}\left(r(m' - 1) - m\right) = 0 \label{5a}
\eeq
 ostensibly nonlinear in $\Phi$ may be reduced to the linear form
\beq
y'' -\frac{1}{r^2} \left(r(m'-1)-m\right) y'=0 \label{5b}
\eeq
by making the change of variables $e^{2\Phi(r)} = y^2(r)$. Equation (\ref{5b}) may be solved explicitly by
\beq
y=c_1 \int \exp \left(\int{\left(\frac{m'-1}{r} - \frac{m}{r^2}\right)dr}\right) + c_2 \label{5c}
\eeq
where $c_1$ and $c_2$ are integration constants that may be settled by considering the boundary conditions. For stellar distributions $\Lambda$ is ignored and so (\ref{5c}) provides an algorithm to detect all static isotropic  perfect fluid  solutions in (2+1) dimensions. Once a suitable form for $m(r)$ is selected, $\Phi$ can be determined (theoretically) and hence the density and pressure may be obtained to complete the model. But we shall not pursue these ideas here as we shall require anisotropic particle pressure for our model.

We are going to solve the resulting Einstein's equations, for static spherically symmetric perfect fluids in (2+1) dimensions, with a maximally localized source of energy having
the minimal width, Gaussian, mass/energy distribution
\begin{equation}
\rho=\frac{M}{4\pi\alpha}~  \text{exp}\left(-\frac{r^2}{4\alpha}\right),
\end{equation}
where $M$ is the total mass of the source. This is due to the coordinate coherent states approach to noncommutative
geometry with the noncommutative parameter $\theta$ is a small ($\sim$ Plank length$^2$) positive number.

By solving the Einstein equations with an EOS $p_r= - \rho$, as a matter source, we have the following
relationship (see Ref. \cite{{Rahaman(2013)}})

\begin{equation}
e^{2\Phi}= -A+2Me^{-\frac{r^2}{4\alpha}}-\Lambda r^2,
\end{equation}
where $A$ is an integration constant. In the limit $\frac{r}{\sqrt{\alpha}}\rightarrow ~\infty$, Eq. (11),
is reduced to the BTZ black hole where the  constant term $A$ plays the role of the
mass of the BTZ black hole, i.e., $A = M$.

In order to proceed with our investigation, we choose a specific  mass function $m(r)$, for closing the system.
For this purpose, we are now interested in the noncommutative geometry inspired mass function (see Refs \cite{{Spallucci},{Garattini}})
into the following form:
\begin{equation}
\tilde{m}=\frac{M}{\pi ^{(\tilde{m}-2)/2}}\gamma \left[\frac{\tilde{m}}{2}, \chi^2 \left(\frac{r}{2M}\right)^2\right],
\end{equation}
where $\chi^2= M^2/\alpha$ and $\gamma \left(\frac{a}{b}; x\right)$ is the Euler lower Gamma function defined by
\begin{equation}
\gamma \left(\frac{a}{b}; x\right) \equiv \int^x_0 u^{a/b} e^{-u}\frac{du}{u}.
\end{equation}
For a BTZ black hole, $\tilde{m}=2$, we obtain the following expression for the mass function as
\begin{equation}
m(r)= M\int^{r^2/4\theta}_0 e^{-t}dt= M \left[1-\text{exp}\left(-\frac{r^2}{4\alpha}\right)\right],
\end{equation}
At the origin, $m(0) = 0$, which is consistent with the solution of Eq. (12) and we notice that the
parameter $\chi$ plays a critical role in determining the horizons.

An interesting feature of the solution is the horizon. Corresponding to $A= M$ given in Eq. (11),
and letting the function $g_{tt}(r_h) =0$, gives the event horizon(s) which is depicted in Fig. 1 for
different values of $\chi$. Thus we find three possible cases \cite{Rahaman(2013)} :\\
(i) for $M>$ $M_0=0.214 \sqrt{\alpha}$ there are two horizons i.e., when $\chi > 0.214 $; \\
(ii) for $M =$ $M_0 = 0.214 \sqrt{\alpha}$ with one degenerate horizon i.e., when $\chi = 0.214 $; \\
(iii) for 0$<$ $M <$ $M_0$ no horizon for $\chi < 0.214 $.\\
where $M_0$ is the existence of a lower bound for a black hole mass
and represents its final state at the end of Hawking evaporation process. It's also
 clear From Fig. 1 that below the minimal mass there is no black hole exists.

\begin{figure}[ptb]
\begin{center}
\vspace{0.3cm}
\includegraphics[width=0.55\textwidth,natwidth=610,natheight=642]{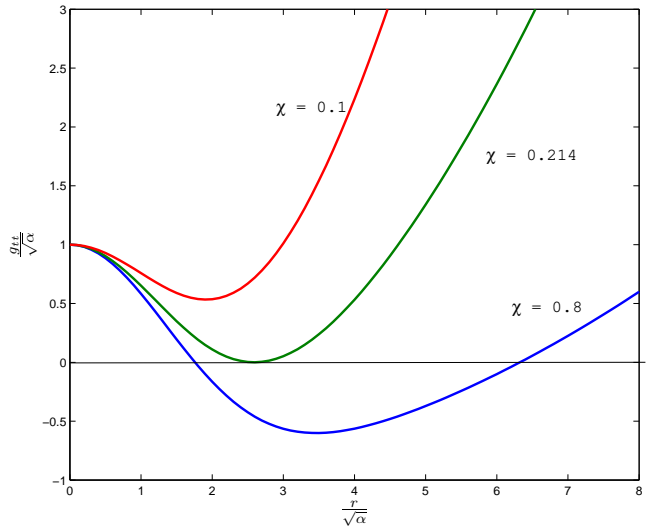}
\end{center}
\caption{The function $g_{tt}$ cuts the r-axis gives the event horizons for different
values of $\chi$ = 0.8, $\chi$ = 0.214 and $\chi$ = 0.1, respectively. }
\end{figure}

\section{Matching at junction interface and surface stresses}

For the specific gravastar model we match the interior gravastar geometry, given in Eq. (1),
with an exterior geometry associated with the BTZ solution
\begin{equation}
ds ^2 = - \left(-M-\Lambda r^2\right) dt^2 + \left(-M-\Lambda r^2\right)^{-1} dr^2 + r^2 d\theta^2,
\end{equation}
both interior and exterior matched at the junction surface $\Sigma$,
situated outside the event horizon, $a > r_h$. As the gravastar solution
does not possess a singularity at the origin and has no event horizon,  we are
interested in the case  $\chi < 0.214 $ and there is no event horizon yielding a solution.

Since, the outer solutions have a zero stress-energy, while at the junction surface $\Sigma$,
both will have a non-zero stress-energy. The junction hypersurface is a timelike hypersurface
defined by the parametric equation $f(x^{\mu}(\xi^i)) = 0$, where $\xi^i$ = ($\tau, ~\theta$) represents the intrinsic
coordinates on the hypersurface and $\tau$ is the proper time, respectively.

In order to proceed one can write the line element for
intrinsic metric to the $\Sigma$, as
\begin{equation}
ds ^2_{\Sigma} = - d\tau^2 + a^2 d\theta^2.
\end{equation}
For the purpose of this paper we matched our interior geometry by the exterior BTZ solution, the three velocity
of a piece of stress energy at the junction surface is given by: $\xi^{\mu} (\tau, \theta)$ = $(t(\tau), a(\tau), \theta)$
\begin{equation}
U^{\mu}_{\pm} = \left(\frac{dt}{d\tau}, \frac{da}{d\tau}, 0\right)=
\left( \frac{\sqrt{1-\frac{2m_{\pm}}{a}+\dot{a}^2}}{1-\frac{2m_{\pm}}{a}}, \dot{a},0\right),
\end{equation}
where the $(\pm)$ correspond to the exterior and interior spacetimes, with $m_{\pm}$ are defined as
interior and exterior mass, respectively.

Also the normal unit vector ($n^{\pm}_{\mu}$) to the boundary can be  defined as ( with $n^{\mu}n_{\mu}= 1$ and $U^{\mu}n_{\mu}=0$)
\begin{equation}
n^{\pm}_{\mu} = \left(- \dot{a},\frac{\sqrt{1-\frac{2m_{\pm}}{a}+\dot{a}^2}}{1-\frac{2m_{\pm}}{a}}, 0\right).
\end{equation}
 At the junction surface the components of the extrinsic curvature tensor reads
\begin{equation}
K^{\pm}_{ij} = -\eta_{\nu}\left(\frac{\partial^2 x^{\nu}}{\partial\xi^i \partial\xi^j}+\Gamma^{\nu\pm}_{\alpha\beta}\frac{\partial x^{\alpha}}{\partial\xi^{i}}\frac{\partial x^{\beta}}{\partial\xi^{j}}\right),
\end{equation}
where $\xi^i$ = ($\tau, ~\theta$), represent the coordinates on the shell. Here, in general $K_{ij}$ is
discontinuous at the junction surface, the discontinuity in the second fundamental forms is defined as
\begin{equation}
\mathcal{K} _{ij} =   K^+_{ij}-K^-_{ij}.
 \end{equation}
Then using Lanczos equation the Einstein equations lead to the following form
 \begin{equation}
\mathcal{S}^{i} _{~j} = -\frac{1}{8\pi} \left(\mathcal{K}^{i} _{~j}-\delta ^{i} _{~j}\mathcal{K}^{k} _{~k} \right),
 \end{equation}
where $\mathcal{S}^{i} _{~j}$ is the surface stress-energy tensor on $\Sigma$, with discontinuity of the
extrinsic curvature defined by $\mathcal{K}^{i} _{~j}$.

Now, let us calculate the non-trivial components of the extrinsic curvature
for the interior spacetime (1), and the exterior BTZ solution (15), are given by
\begin{equation}
K^{\tau +} _{~\tau} =  \frac{-\Lambda a+\ddot{a}}{\sqrt{-M-\Lambda a^2+\dot{a}^2}},
 \end{equation}
\begin{equation}
K^{\tau -} _{~\tau} =  \frac{\frac{m}{a^2}-\frac{m'}{a}+\ddot{a}}{\sqrt{1-\frac{2m(a)}{a}+\dot{a}^2}},
 \end{equation}
 and
\begin{equation}
K^{\theta +} _{~\theta} = \frac{1}{a} {\sqrt{-M-\Lambda a^2+\dot{a}^2}},
 \end{equation}
\begin{equation}
K^{\theta -} _{~\theta} = \frac{1}{a} {\sqrt{1-\frac{2m(a)}{a}+\dot{a}^2}},
 \end{equation}
where the prime denotes a derivative with respect to $r$ and  dot stands for $d/d\tau$. Therefore, the stress-energy tensor
(21) in the most general form of the surface energy density $\sigma$, and the surface pressure, $\mathcal{P}$, as
$\mathcal{S}^{i} _{~j}$ = diag $\left( -\sigma, \mathcal{P}\right)$.

After some algebraic manipulation and using Lanczos equation, we obtain the
energy density and the surface pressures are given by
\begin{equation}
\sigma= -\frac{1}{8\pi a}\left({\sqrt{-M-\Lambda a^2+\dot{a}^2}}- {\sqrt{1-\frac{2m(a)}{a}+\dot{a}^2}}\right),
 \end{equation}
\begin{equation}
\mathcal{P} = \frac{1}{8\pi a}\left[\frac{-M-2\Lambda a^2+\dot{a}^2+\ddot{a}}{\sqrt{-M-\Lambda a^2+\dot{a}^2}}-
\frac{1-\frac{m}{a}-m'+\dot{a}^2+\ddot{a}}{\sqrt{1-\frac{2m(a)}{a}+\dot{a}^2}}\right].
 \end{equation}
Note that by definition the surface tension $\sigma$ has the opposite sign as of the surface pressure $\mathcal{P}$.
Now, we shall also use the conservation identity in the form $\mathcal{S}^{i} _{~j|i}$ =$[T_{\mu\nu}e^{\mu}_{j}n^{\nu}]^{+}_{-}$,
where $[X]^{+}_{-}$ represents the discontinuity across the surface interface. The method is  developed  in Ref \cite{Lobo}.
To study the stability of the solutions under perturbations we encroach on the momentum flux term
$F_{\mu}=T_{\mu\nu}U^{\nu}$ in the right hand side corresponds to the net discontinuity. With the definitions
of conservation identity one can  convert this into conserved energy and momentum of the surface stresses at the junction
interface.

It is useful to introduce the conservation equation in a form that relates the surface energy and surface pressure
with the work done by the pressure and the energy flux on the shell given by the
equation $\mathcal{S}^{i} _{~\tau |i}$ = -$\left[\dot{\sigma}+\frac{\dot{a}}{a}(\sigma+\mathcal{P})\right]$.
Then the conservation identity provides the following relationship
\begin{equation}
\sigma^{'} = -\frac{1}{a}\left(\sigma+\mathcal{P}\right),
 \end{equation}
where $\sigma^{'}= \frac{\dot{\sigma}}{\dot{a}}$. Now taking into account the Eqs. (26-27), the
Eq. (28) has the form
\begin{equation}
\sigma^{'} = \frac{1}{8\pi a^2}\left[\frac{-M+\dot{a}^2-\ddot{a}}{\sqrt{-M-\Lambda a^2 +\dot{a}^2}}-
\frac{1-\frac{3m}{a}+m^{\prime}+\dot{a}^2-\ddot{a}}{\sqrt{1-\frac{2m}{a} +\dot{a}^2}}\right],
 \end{equation}
and at the static solution $a_0$ which reduce to
\begin{equation}
\sigma^{'}(a_0) = \frac{1}{8\pi a_0^2}\left[\frac{-M}{\sqrt{-M-\Lambda a_0^2}}-
\frac{1-\frac{3m}{a_0}+m^{\prime}(a_0)}{\sqrt{1-\frac{2m}{a_0}}}\right],
 \end{equation}
which play a crucial role in determining the stability regions
as we considered below.

\section{Stability analysis}
In this section, we investigate the stability of gravastar in a perturbative treatment of the shell dynamics,
more precisely, the linearized stability of the solutions.
For that we rearrange the Eq. (26) to obtain the thin-shell equation of motion
\begin{equation}
\dot{a}^{2}+V(a) =0,
 \end{equation}
where the potential $V (a)$ is given by
\begin{equation}
V(a) =\frac{\mathcal{G}_{1}(a)+\mathcal{G}_{2}(a)}{2}-\left[\frac{\mathcal{G}_{1}(a)-\mathcal{G}_{2}(a)}{16\pi a \sigma (a)}\right]^2-\left[4\pi a \sigma (a)\right]^2,
 \end{equation}
where, for notational convention we have used  $\mathcal{G}_{1}(a)= {\sqrt{1-\frac{2m(a)}{a}+\dot{a}^2}}$ and
$\mathcal{G}_{2}(a)={\sqrt{-M-\Lambda a^2+\dot{a}^2}}$, respectively.

 Now using the surface mass of the thin shell $m_s = 2\pi a \sigma$, allows to write the potential in the form
\begin{equation}
V(a) = \mathcal{S}-\left(\frac{\mathcal{T}}{4m_s}\right)^2-(2m_s)^2,
 \end{equation}
with
\begin{equation}
\mathcal{S}=\frac{\mathcal{G}_{1}(a)+\mathcal{G}_{2}(a)}{2}~~ \text{and} ~~\mathcal{T}= \frac{\mathcal{G}_{1}(a)-\mathcal{G}_{2}(a)}{2}.
 \end{equation}

For the stability analysis of static solutions at $a_0$ under the radial perturbations,
we consider the Taylor expansion of the potential function $V(a)$  around $a_0$ up to second order, given by
\begin{eqnarray}
V(a) &=&  V(a_0) + V^\prime(a_0) ( a - a_0) +
\frac{1}{2} V^{\prime\prime}(a_0) ( a - a_0)^2  \nonumber \\
&\;& + O\left[( a - a_0)^3\right],
\end{eqnarray}
where the prime denotes the derivative with respect to $a$. The existence and stability of the static solutions
depend upon the inequalities that $V(a_0)$ has local
minimum at $a_0$ and $V^{\prime\prime}(a_0)>0$. The first derivative of the potential is given by
 \begin{equation}
V^{\prime}(a) =\mathcal{S}^{\prime}-8m_s m_s^{\prime}-\frac{\mathcal{T}}{8m_s}\left(\frac{\mathcal{T}}{m_s}\right)^{\prime},
 \end{equation}
using the conditions  $V^{\prime}(a_0) =0$, we can write the equilibrium relationship as
 \begin{equation}
\bm{X}\equiv m_s^{\prime}=\frac{1}{8m_s}\left[\mathcal{S}^{\prime}-\frac{\mathcal{T}}{8m_s}\left(\frac{\mathcal{T}}{m_s}\right)^{\prime}\right]
 \end{equation}

Finally, second derivative of the potential gives
 \begin{equation}
V^{\prime \prime}(a) =\mathcal{S}^{\prime \prime}-8m_s m_s^{\prime \prime}-8 m_s^{\prime 2}-\frac{1}{8}
\left[\left(\frac{\mathcal{T}}{m_s}\right)\left(\frac{\mathcal{T}}{m_s}\right)^{\prime \prime}+\left(\frac{\mathcal{T}}{m_s}\right)^{\prime 2}\right].
 \end{equation}

To evaluate for a static equilibrium configuration for the stability, we rewrite
the conservation of the surface stress-energy tensor, as $a\sigma^{\prime} = -\left(\sigma+\mathcal{P} \right)$, and
taking into account the new parameter  $\eta=\frac{\mathcal{P}^{\prime}}{\sigma^{\prime}}$, the surface mass of the thin shell
given by
\begin{equation}
m_s^{\prime \prime} =\frac{2\pi}{a}\left(\sigma+\mathcal{P}\right)\eta.
\end{equation}
Here the parameter $\eta$, which is interpreted as the subluminal speed of sound, has been used to present
the stability regions without using the surface equation of state.

Now, evaluated for a static equilibrium configuration for the stability and taking into account Eq. (38),
with $V^{\prime \prime}(a_0) > 0$, we have
\begin{equation}
\eta_{0}\frac{d\sigma^{2}}{da} \Bigl\rvert_{a=a_0} > \Pi,
\end{equation}
by using Eq. (39), where $\eta_{0}$ = $\eta(a_0)$ and $\Pi$, for notational simplicity, we define
a simply behaving function of the form
\begin{equation}
 \Pi \equiv \frac{1}{2\pi^2 a_0}\left(\bm{X^2-\bm{Y}}\right),
\end{equation}

\begin{figure}[ptb]
\begin{center}
\vspace{0.3cm}
\includegraphics[width=0.4\textwidth,natwidth=610,natheight=642]{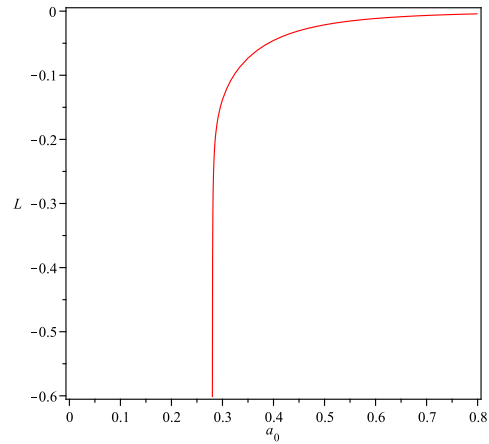}
\end{center}
\caption{In these figure we have plotted the dimensionless parameter $L= d{\sigma^2}/ da |_{a_0}$
corresponding to the value of $\chi$ = 0.16 with $\Lambda \sqrt{\alpha}$ = - 0.02.
The figure is shown for the specific case when m(a) $<$ M/2. }
\end{figure}

\begin{figure}[ptb]
\begin{center}
\vspace{0.3cm}
\includegraphics[width=0.4\textwidth,natwidth=610,natheight=642]{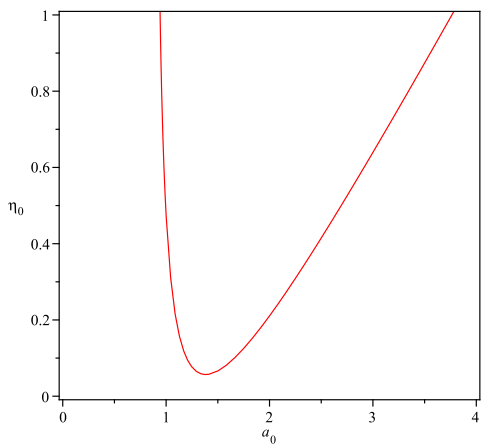}
\end{center}
\caption{The stability region is shown by the plots for $\chi$ = 0.16 and $\Lambda \sqrt{\alpha}$ = - 0.02.
The stability region is depicted below the surfaces which is sufficient close to the event horizon. }
\end{figure}
where
\begin{equation}
\bm{Y}=\frac{\mathcal{S}^{\prime \prime}}{8}-\frac{1}{64}
\left[\left(\frac{\mathcal{T}}{m_s}\right)\left(\frac{\mathcal{T}}{m_s}\right)^{\prime \prime}+\left(\frac{\mathcal{T}}{m_s}\right)^{\prime 2}\right]
\end{equation}

In order to analyze the stable equilibrium regions of the solution we adopt the the following inequalities
\begin{equation}
\eta_{0} > \Im,~~~ \text{if}~~~ \frac{d\sigma^{2}}{da}\Bigl\rvert_{a=a_0}>~0,
\end{equation}
\begin{equation}
\eta_{0} < \Im,~~~ \text{if}~~~ \frac{d\sigma^{2}}{da}\Bigl\rvert_{a=a_0}<~0,
\end{equation}
with the definition
\begin{equation}
\Im \equiv \Pi \left(\frac{d\sigma^{2}}{da}\Bigl\rvert_{a=a_0}\right)^{-1}.
\end{equation}

\section{Region of stability}\noindent
We shall in this section consider the static solution for the stability analysis and
deduce the stability region by considering the inequalities Eqs. (43-44).
For this purpose we shall impose a positive surface energy density $\sigma >0$, which
indicates $m(a) < M$. For the case of m(a) $< M$, and using the condition $\chi < 0.0214$
one can prove that $d\sigma^{2}/da\Bigl\rvert_{a_0} < 0$. Therefore the stability
region is constrained by the inequality (44). To justify our assumption we use graphical representation
due to complexity of the expression $\Im$, which is plotted in Fig. 2
for the specific case of m(a) = M/2 and when $\chi$ = 0.16 with $\Lambda \sqrt{\theta} = - 0.02$.
From Fig. 2 it clear the the solution does not corresponds any horizon.

In order to study the stability region we use the graphical representation (Fig. 3) for the
case when $\chi$ = 0.16. We have examined the stability of the model based on the
speed of sound which should lies within the limit (0, 1]. According to Fig. 3, that the above
stability regions is sufficiently closed to the event horizon which decreases for increasing $a$,
and increases again as $a$ increases.

\section{Conclusions}\noindent
  In this paper, we have studied the stability of gravastar solution in a (2+1)-dimensional anti-de Sitter space
given in a context of noncommutative geometry. At first we derive BTZ solution assuming the source of energy
density as point-like structures in favor of smeared objects, where the particle mass is diffused
throughout a region of linear size $\sqrt{\alpha}$ and is described by a Gaussian function of finite width
rather than a Dirac delta function. In Fig. 1, it was shown that depending on the values of
$\chi$ the metric displays different causal structure : existence of two horizons,
one horizon or no horizons.

To search for gravastar solution we matched the interior geometry for specific
mass function, with an exterior BTZ solution at a junction interface situated outside the event horizon.
However, to obtain a realistic picture, we explored the linearized stability analysis of the
surface layer which is sufficient close to the event horizon. Considering the static solution
to find the stability  region we use the graphical representation (Fig. 3) based on speed
of sound which lies within the limit (0, 1]. At this point we would like to discuss that large
stability region do exists and sufficiently closed to the event horizon. Therefore, considering the
model one would state that it is difficult to distinguish the exterior geometry of the gravastar
from a black hole.

\section{Acknowledgments}\noindent
We thank Dr. Douglas Singleton and Dr. Farook Rahaman for useful
discussions. AB would like to thank the authorities of the Inter-University Centre for
Astronomy and Astrophysics, Pune, India for providing the Visiting Associateship.

\end{document}